\documentclass[mathleft
]{an}
\usepackage{graphicx}
\usepackage{times}
\begin{document}

\Pagespan{789}{}
\Yearpublication{2006}%
\Yearsubmission{2005}%
\Month{11}%
\Volume{999}%
\Issue{88}%

\title{A possible link between high rotation measure and CSS-GPS sources.}

\author{Alice Pasetto \inst{1}\fnmsep\thanks{
  \email{apasetto@mpifr-bonn.mpg.de}\newline}
   		Alex Kraus \inst{1}, 
		Karl-Heinz Mack \inst{2},
          	Gabriele Bruni \inst{1,}\inst{2},
		Carlos Carrasco-Gonz\'alez \inst{3}
}
\titlerunning{Possible link between high-RM and CSS-GPS}
\authorrunning{Pasetto et al.}
\institute{Max-Planck-Institut f\"ur Radioastronomie (MPIfR), 
              Auf dem H\"ugel 69, 53121 Bonn, Germany
              \and
             Istituto di Radioastronomia (IRA-INAF), Via Gobetti 101, 40129 Bologna, Italy
             \and
             Instituto de Radioastronom\'ia y Astrof\'isica (IRAf-UNAM), Antigua Carretera a P\'atzcuaro 8701, 58089 Morelia, Michoac\'an, M\'exico
             }

\received{}
\accepted{}
\publonline{}

\keywords{galaxies: active -- radio continuum: galaxies -- polarisation  }

\abstract{%
We report here the study, in the radio band, of a sample of candidates of high Rotation Measure (RM). The point-like objects (at kpc scale) were selected by choosing unpolarized sources from the NVSS which show significant linear polarization at 10.45 GHz. Assuming in-band depolarization, this feature suggests the presence of a very dense medium surrounding them in a combination of a strong magnetic field. Further single-dish observations were performed with the 100-m Effelsberg telescope to characterise the SEDs of the sample and to well determine their RM in the 11 to 2 cm wavelength range. Besides, a wideband (L, S, C and X band ) full polarisation observational campaign was performed at the JVLA facility. It allows us to analyse the in-band RM for the most extreme objects. Some Effelsberg results and analysis, and preliminary JVLA results are presented. 
The observations reveal that sources with young, newly growing, radio components at high frequency (i.e. GPS and HFP sources) are characterised by a really dense and/or a magnetised medium that strongly rotates the polarisation angle at the different frequencies, leading to a high-RM.
   }
\maketitle

\section{Introduction}
The unification scheme for radio loud AGN, after which their appearance depend strongly on their orientation (\cite{OB82}, \cite{UP95}), is by now accepted by the most of the scientific community, but still several open questions remain. Some of them concern the environment, i.e. the composition of the ambient medium and/or the strength of the magnetic field. Some others regard the status of an AGN, i.e. if it is characterised by some periodic activity phase (\cite{Ma06, Saikia09, Cz09}) or if it is experiencing some evolution from young quasar, like the High Frequency Peakers (HFP, \cite{Dcasa00}), the Giga-Hertz Peaked-spectrum Sources (GPS, \cite{Gopal-Krishna83} \cite{O'Dea91}) and the Compact Steep-spectrum Sources (CSS, \cite{Saikia88}, \cite{Fanti90}), to large-scale radio sources (\cite{O'Dea98}).
The study of the Faraday rotation and depolarization of radio sources is a powerful tool that can probe the interstellar medium and the strength of the magnetic field of the host galaxy. 

The Faraday rotation occurs when an electromagnetic wave passes through a magnetised plasma. For an homogeneous medium, it depends linearly with $\lambda^{2}$ and it is described by the Rotation Measure (RM) which can be expressed as (\cite{Burn66}): 
\[RM = 0.81 \int n_{e} B_{\parallel}dl \quad [rad/m^2],  \]
\noindent
where n$_{e}$, is the electron density of the medium, B$_{\parallel}$ is the component of the magnetic field along the line of sight and \textit{l} is the geometrical depth of the medium along the line of sight.
If the medium is inhomogeneous or unresolved, the RM can change within the source and a deviation from the $\lambda^{2}$ law occurs (\cite{Burn66}, \cite{Val80}, \cite{SaSa88}). This can be an indication of multiple RM components.

When the Faraday rotation causes the reduction of the fractional polarization, the source is subject to depolarization, the behaviour of the fractional polarization (\textit{m}) with $\lambda^{2}$. Several depolarization and repolarization (where the polarized fraction increasing at longer wavelengths) models have been developed (\cite{Burn66}; \cite{Tribble91}; \cite{Homan02}; \cite{Fanti04}; \cite{Rossetti08}; \cite{RosMan09}; \cite{Man09}; \cite{Hovatta12}).

Some observational works have found sources with very high-RM with single dish, local interferometer and also with higher resolution VLBI technique (\cite{Kato87}, \cite{Zavala04}, \cite{Attridge05}, \cite{Benn05}, \cite{Jorstad07}, \cite{Trippe12}, \cite{Kravchenko15}). However, a deep study of the relationship of the RM and the ambient medium has remained difficult since it requires to study a large sample in a large range of frequencies and with simultaneous observations (to avoid variability effects). 

Here we present preliminary results of an observational campaign performed with the 100-m Effelsberg telescope of a relatively large sample of radio sources. With this project our aim is to study whether any connection is present between the AGN hosting galaxy medium with some evolutionary track and/or some periodic activity phase of the AGN itself. Furthermore, the progress done so far on the interferometric JVLA data on some of the extreme cases in our sample, is presented.

\section{Sample selection criteria}

We selected sources from the NRAO VLA Sky Survey (NVSS; \cite{Condon98}) matching the following criteria: 
\begin{itemize}
\item bright sources with flux density S$_{1.4}$ ${\geq}$ 300 mJy;
\item sources with major and minor axis ${\leq}$45${\arcsec}$;
\item declination $\delta$ ${\geq}$ -10$^{\circ}$;
\item NVSS polarization flux density S$^{pol}_{1.4}$ ${\le}$ 0.87 mJy (i.e. 3$\sigma$$^{pol}_{1.4}$)
\end{itemize}   
\noindent
The last criterium is essential as we are interested in studying sources suffering from a strong in-band depolarization at 1.4 GHz, hint of a possible high-RM.
We cross correlated the obtained sample with the FIRST catalogue (\cite{White97}. In order to increase the probability to select compact and/or high redshift candidates, only unresolved sources were selected ($\leq$5$\arcsec$). The final sample contains 537 bright, point-like and unpolarized sources. 

\section{Observations}
As a first step, we observed the entire sample of 537 sources with the 100-m Effelsberg telescope at 10.45 GHz (FWHM of 69$\arcsec$). The detection of polarisation flux density at 10.45 GHz suggests a strong in-band depolarization at 1.4 GHz probably because of a high-RM. The final sample of high-RM candidates is composed by 30 sources. 
On these, a follow-up programme (at 2.64, 4.85, 8.35, 10.45 and 14.60 GHz) was performed in order to determine their SED and RM value. For the extreme cases, 12 sources having RM $\ge$ 500 rad/m$^2$, JVLA observations in L, S, C and X band were performed. 
Moreover, we also make use of VLBI data, taken with the EVN and the VLBA interferometers, to study the innermost part of these extreme objects.

\section{Effelsberg results and discussion}
\subsection{Spectral index distribution}

Spectral indices between 1.4 GHz (from the FIRST catalogue) and 10.45 GHz were obtained. We compared the spectral index distributions of the 30 high-RM candidates and the unpolarized sources at 10.45 GHz (see Fig.1). While unpolarized sources show a rather symmetric distribution centred at $\alpha$$\simeq$ --0.5, the high-RM candidates shows three different groups of objects:
\begin{itemize}
\item steep spectrum radio sources with $\alpha_{peak}$$\simeq$ --0.8, representative of "lobe dominated", objects where large-scale structures dominate the radio synchrotron emission;
\item flat spectrum radio sources with $\alpha_{peak}$$\simeq$ --0.1, associated to multiple synchrotron peaks due to the superposition of several features (like blazars);
\item inverted spectrum radio sources with $\alpha_{peak}$$\simeq$ +0.2, representative of "core dominated" objects where the dominant synchrotron component is very compact (like GPS sources);
\end{itemize}

Our results suggest that the high-RM candidates are not represented by a particular class of targets. 
\begin{figure}
\begin{center}
\includegraphics[width=0.45\textwidth]{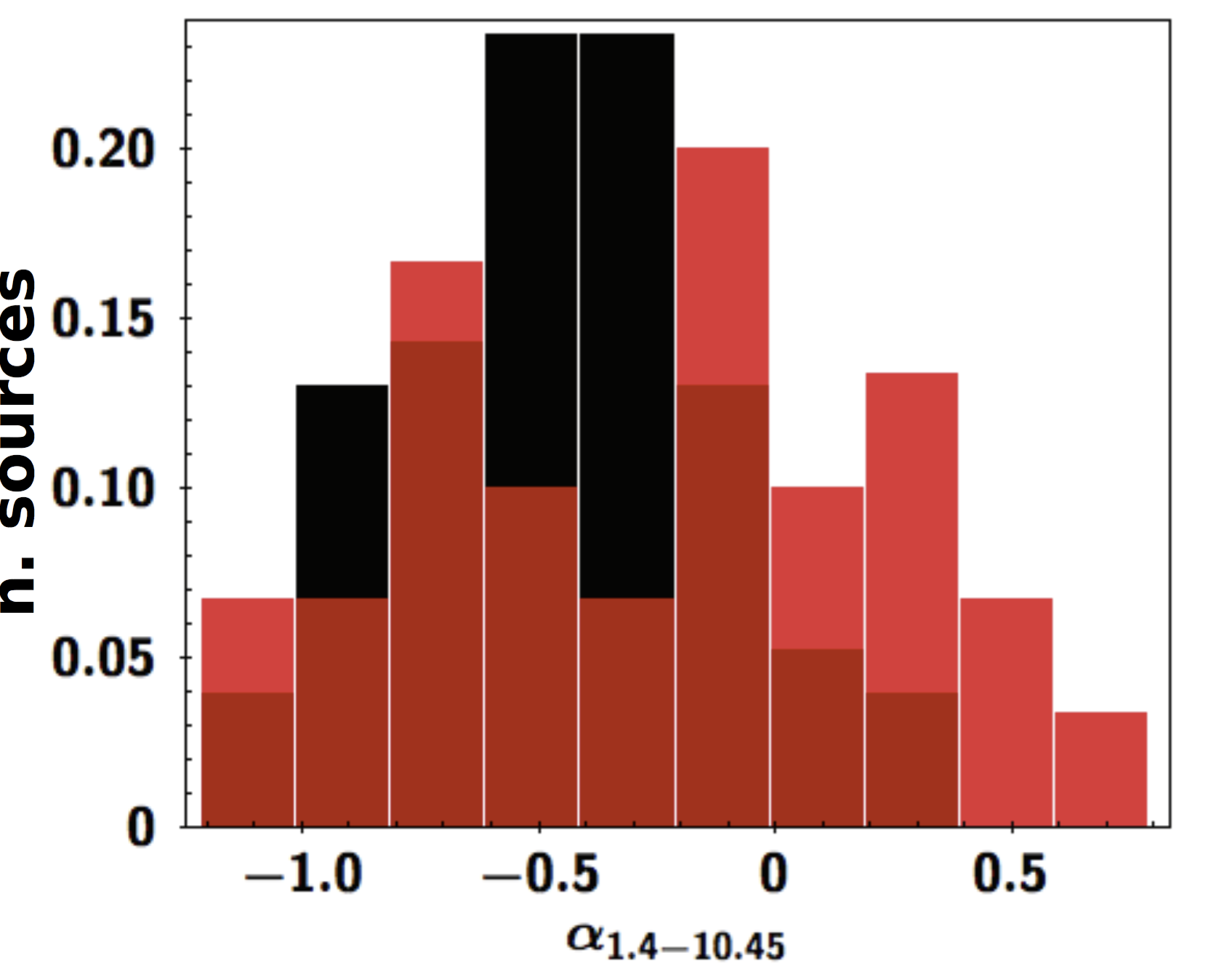}
\caption{Spectral index distributions for the high-RM candidates (30 sources, in red) and the unpolarized sources (77 sources, in black). The histogram has been normalised to the number of sources.}
\end{center}
\end{figure}

\subsection{Spectral energy distributions}
We studied the SEDs of the 30 sources combining the Effelsberg data with data at lower frequencies from literature.
We noticed 3 main groups of SEDs in our sample:
\begin{itemize}
\item $\textit{Old}$: sources with a purely optically thin synchrotron spectrum even at low frequencies.
\item $\textit{GPS-like}$: SEDs with several synchrotron components peaking at frequencies $\ge$ 100 MHz.
\item $\textit{Mixed}$: the combination of the two above.
\end{itemize}
In Fig.2, we show examples of the three object types we identified. Together with the total intensity SED, information on the polarization flux density, the fractional polarization ($\textit{m}$) and the polarization angle ($\textit{$\chi$}$) are presented for each of the targets.
\begin{figure}[!h]
\begin{center}
\includegraphics[width=0.42\textwidth]{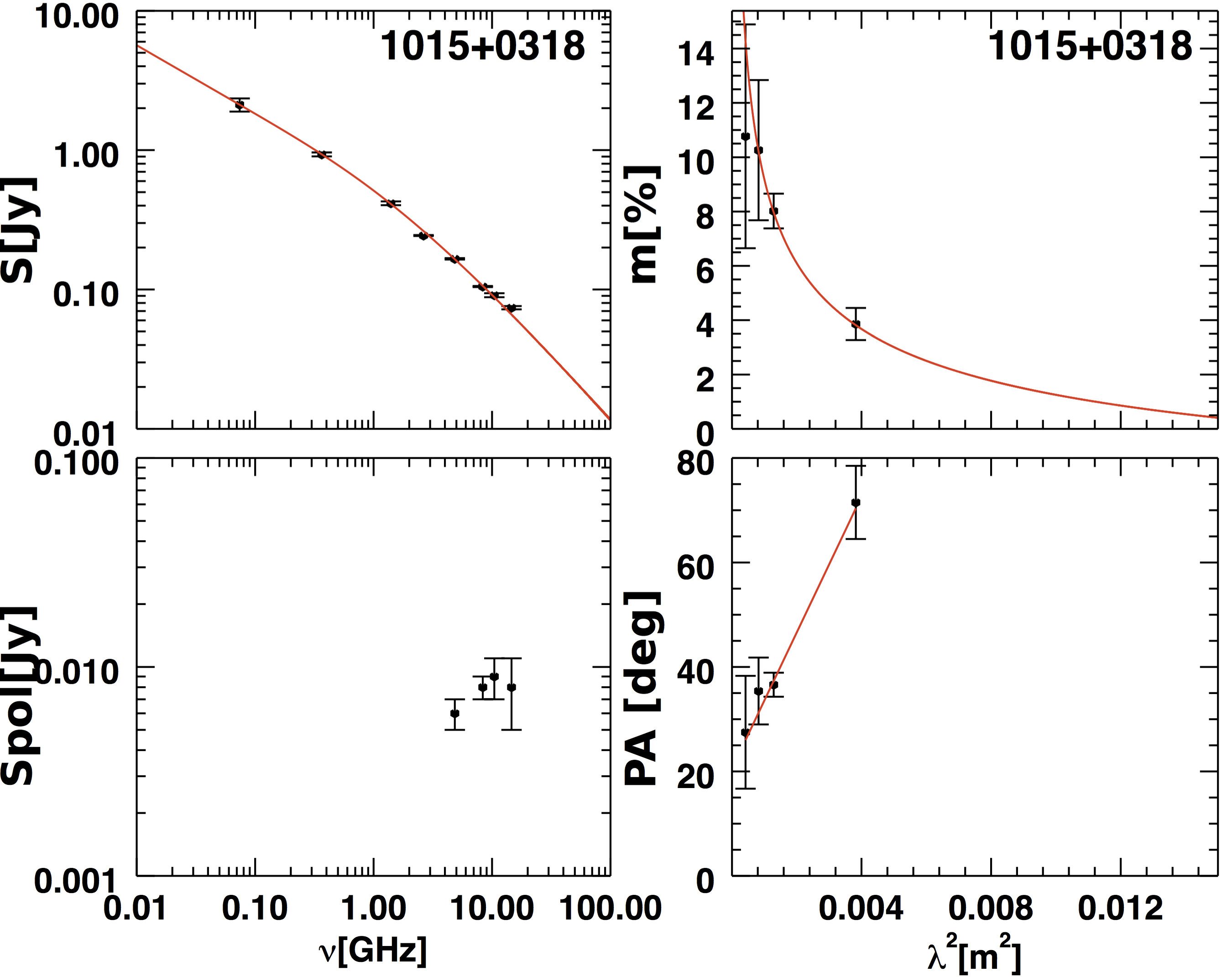}
\includegraphics[width=0.42\textwidth]{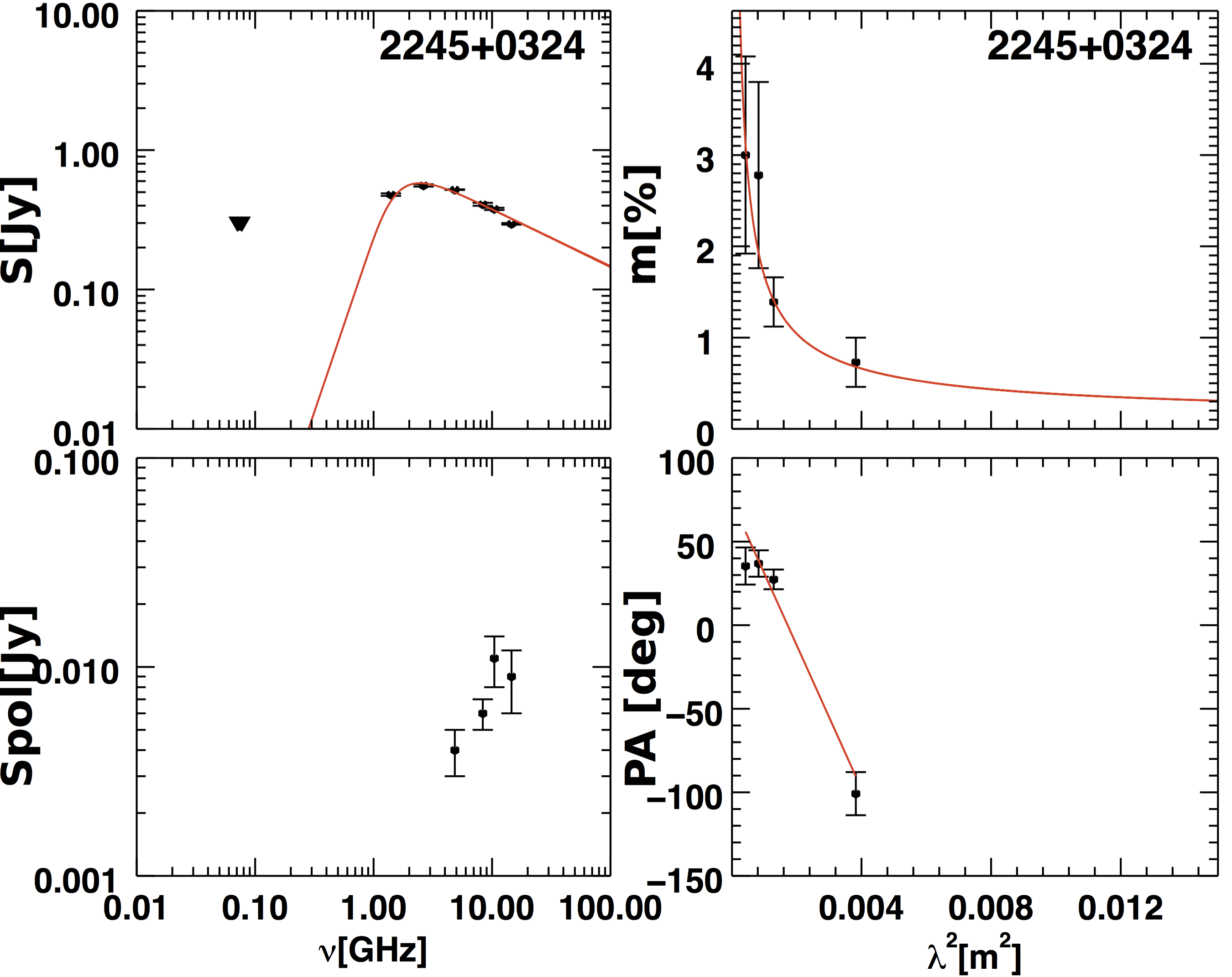}
\includegraphics[width=0.42\textwidth]{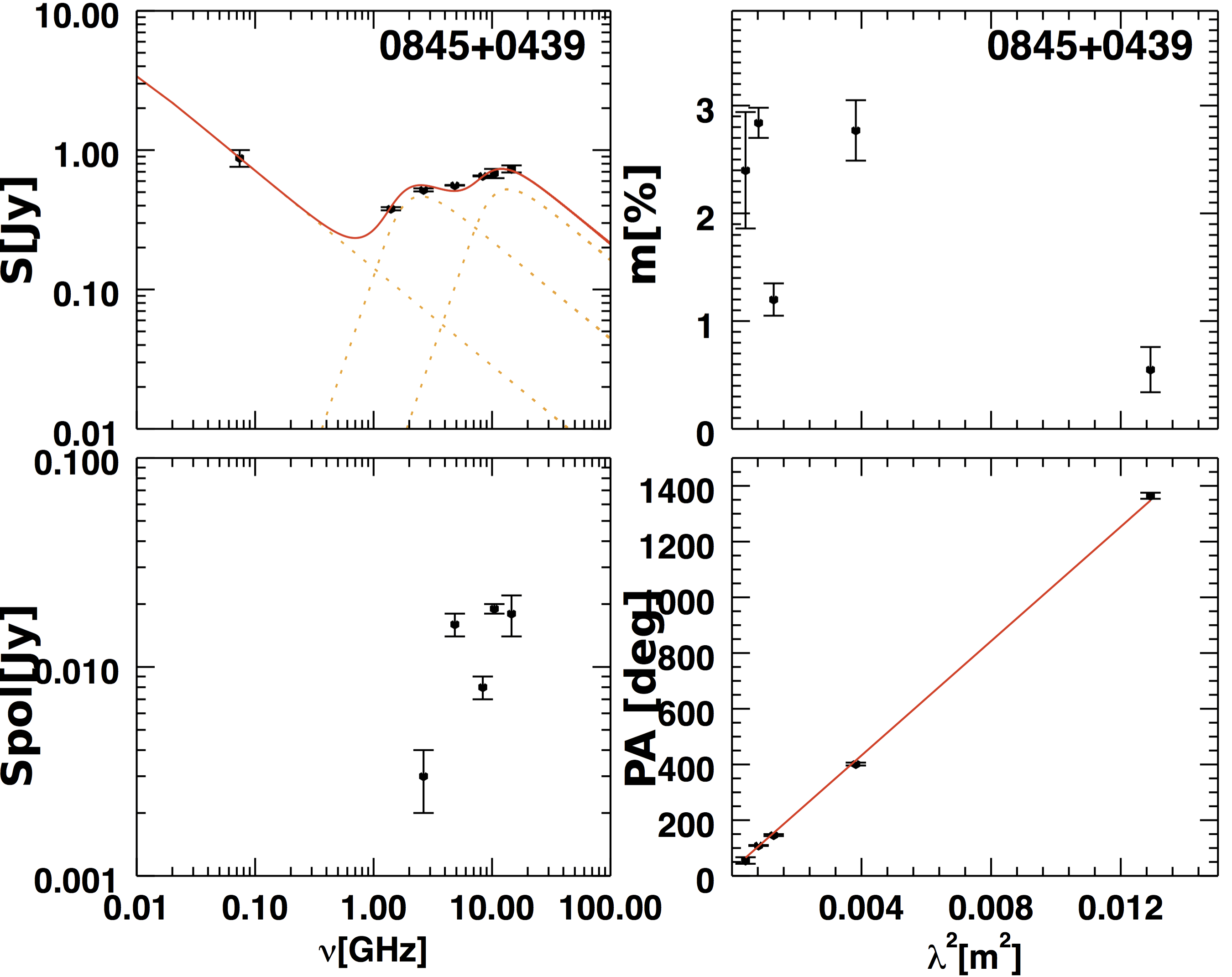}

\caption{SEDs for the sources: 1015+0318 classified as $\textit{Old}$, 2245+0324 classified as $\textit{GPS-like}$ and 0845+0439 classified as $\textit{Mixed}$. Black dot points are the Effelsberg and low frequency data from literature are the black stars; 3-sigmas upper limits are drawn as triangles. Where present, various synchrotron components are plotted with orange dashed line. The fit of the spectra is the red straight line.
Together with the SEDs, polarization information are presented.}

\label{default}
\end{center}
\end{figure}

From the analysis of the radio spectra we find that the targets are equally distributed in the three groups: 1/3 can be considered $\textit{Old}$, thus sources with an extended and probably old synchrotron component, 1/3 can be considered $\textit{GPS-like}$ with a more compact and probably early phases synchrotron components and the remaining can be considered $\textit{Mixed}$, the behaviour of which could be an indication of a restarting radio activity.
Therefore, high-RM sources seem to be associated with compact and/or new growing high frequency components (the $\textit{GPS-like}$ or $\textit{Mixed}$). 

\subsection{Rotation Measure}
We estimate the Rotation Measure (RM) for all the targets as a linear regression fit of the EVPA versus $\lambda$$^{2}$, applying, when necessary, \textit{n$\pi$} ambiguities with a maximum numbers of wraps fixed to 5. 
However we found that 11 sources ($\sim$ 37\% of the sample) deviate from the $\lambda$$^{2}$ law. This suggests that these sources are characterised by several Faraday screens intervening the medium. Indeed, if the radiation passes through different magnetised plasmas, which could rotate differently the polarisation angle, the result would be a non linear regression fit. 
The RM at the source rest frame, RM$_{rf}$, was calculated following the relation:
\[RM_{rf} = RM_{obs} \times (1+z)^{2}\]
In Fig.3 we compare the distributions of the observed $|$RM$|$ of our 30 high-RM candidates and the targets from the Farnes catalogue (\cite{Farnes14}). The two distributions seem to be different. While 80\% of the Farnes targets have a $|$RM$_{obs}$$|$$<$20 rad/m$^2$, 80\% of our targets have a $|$RM$_{obs}$$|$$>$100 rad/m$^2$. In order to check the discrepancy, we also run the Kolmogorov- Smirnov test to the distributions and it gives a probability of 3$\times$10$^{-6}$; the two distributions are different at a confidence level $>$ 95\%.
We also noticed that the RM distribution of the three main object types we have found analysing the SEDs (Fig.4a-b), are different  between them. The $\textit{Mixed}$ targets are those with the largest values for both the $|$RM$_{obs}$$|$ (Fig.4a) and $|$RM$_{rf}$$|$ (Fig.4b). All the $\textit{Mixed}$ targets show $|$RM$_{rf}$$|$ higher than 1000 rad/m$^2$. This seems to suggest that high-RM sources are mainly related to sources with mixed SEDs, i.e. radio sources with a restarting activity.

\begin{figure}[]
\begin{center}
\includegraphics[width=0.45\textwidth]{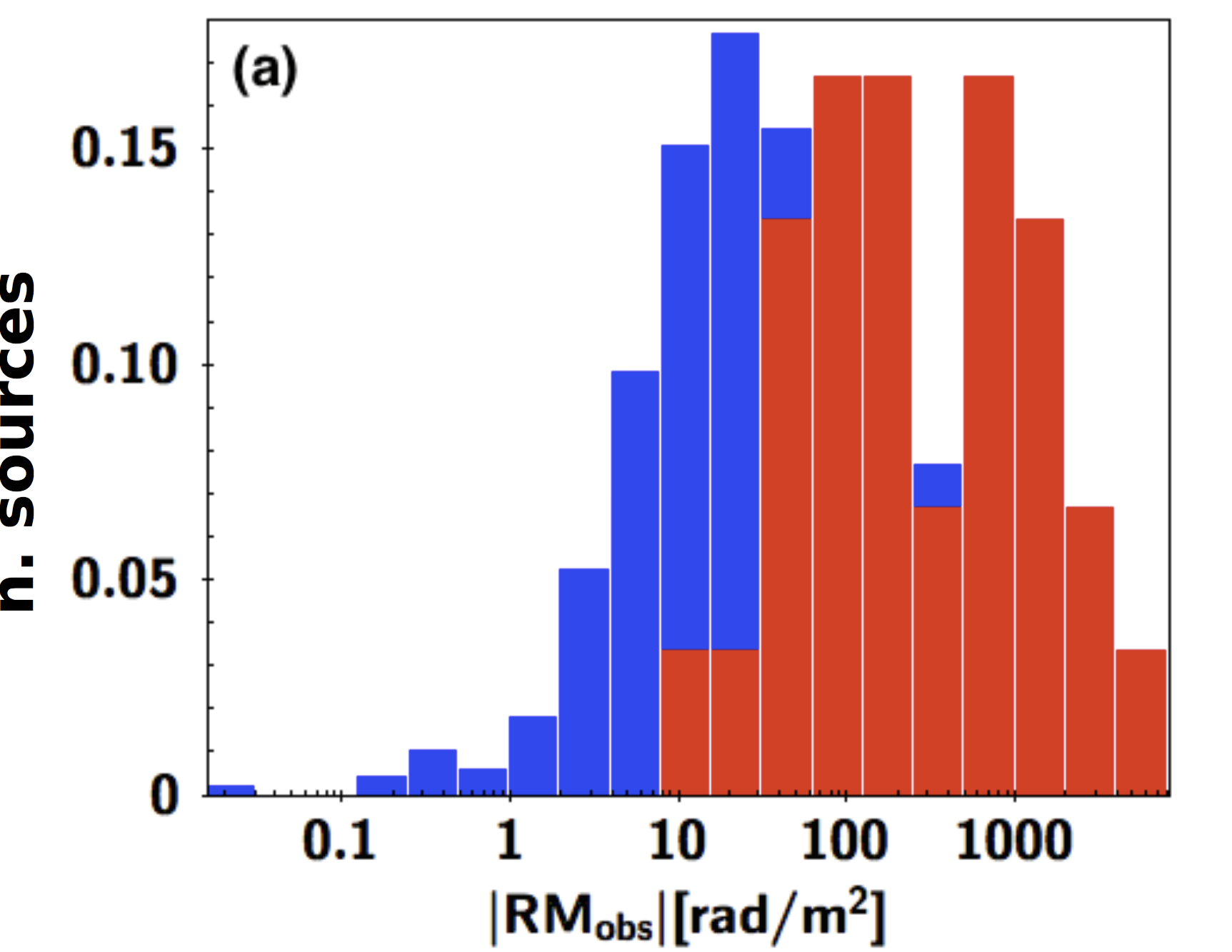}
\caption{The distribution of the fitted rotation measure of our targets (30 sources, red histogram) with the ones from the Farnes catalogue (951 sources, blue histogram).The histogram has been normalised to the number of sources.}
\label{default}
\end{center}
\end{figure}

\begin{figure}[]
\begin{center}
\includegraphics[width=0.4\textwidth]{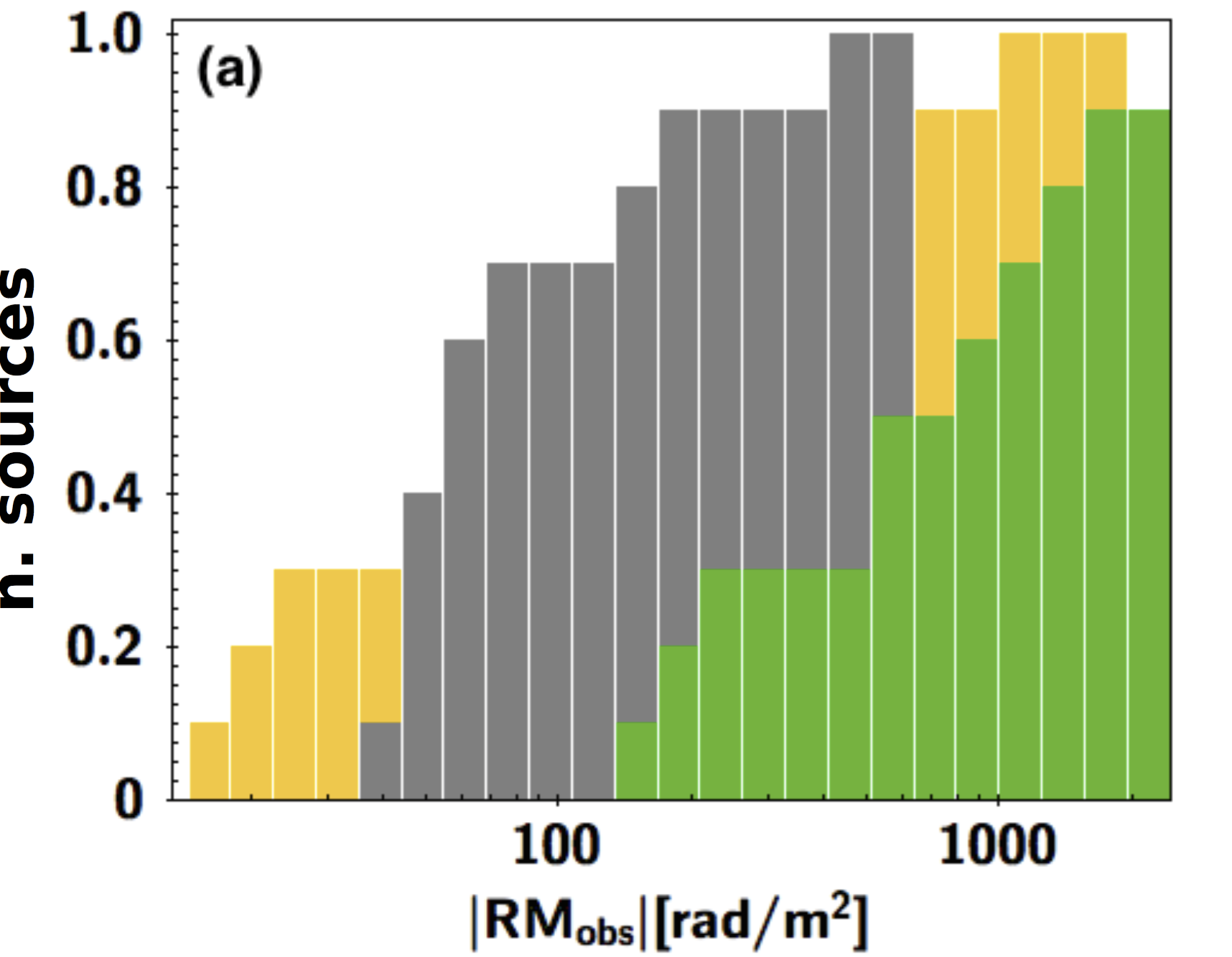}
\includegraphics[width=0.4\textwidth]{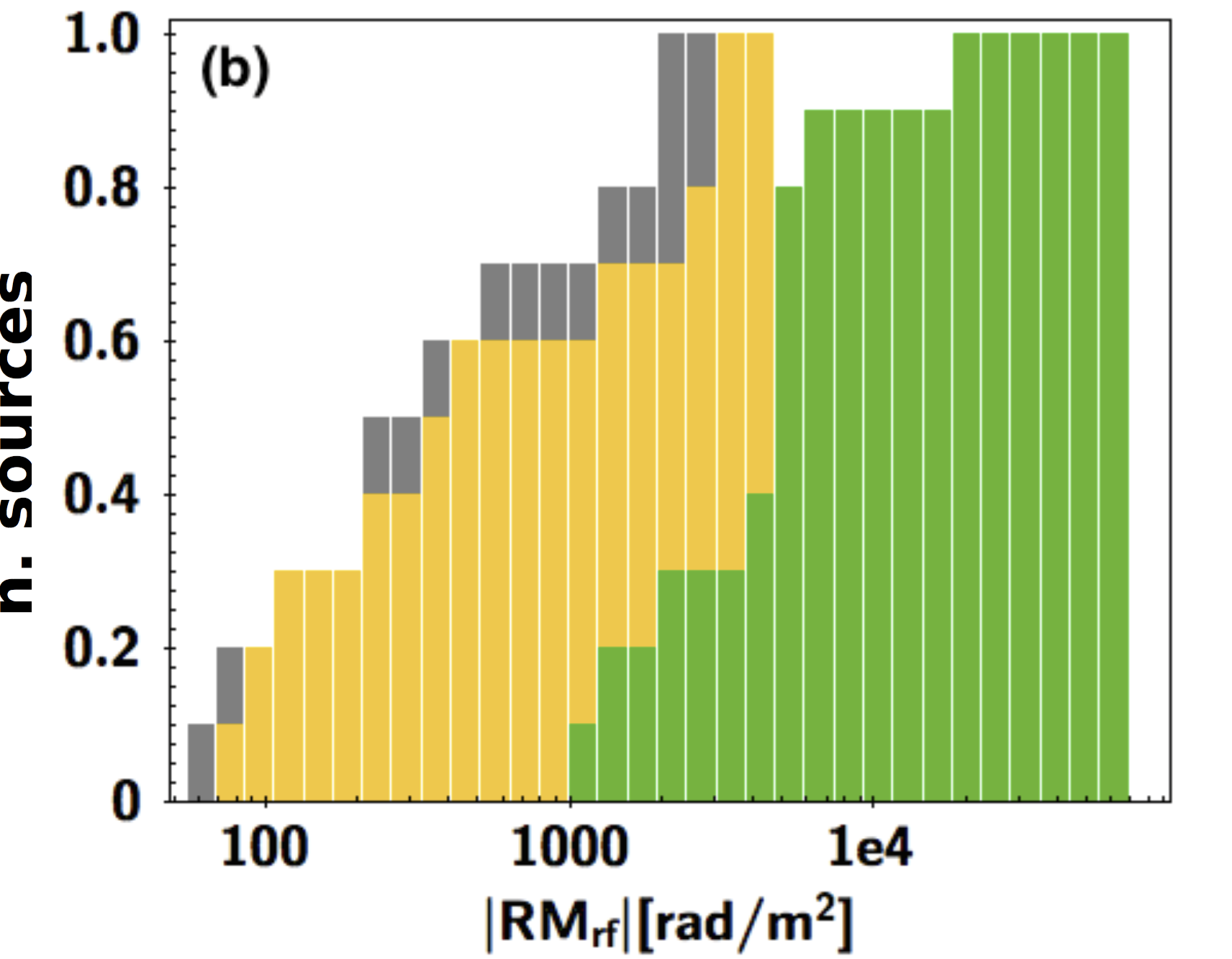}
\caption{\textbf{(a)} cumulative plot of the RM$_{obs}$ for the three object type. Yellow: $\textit{Old}$ type; grey: $\textit{GPS-like}$ type; green: $\textit{Mixed}$ type. \textbf{(b)} cumulative plot of the RM$_{rf}$  for the three objects types. The histograms have been normalised to the number of sources.}
\end{center}
\end{figure}

\subsection{Fractional polarization}
We also analyse the fractional polarization (\textit{m}) as a function of $\lambda^{2}$ in order to study the depolarization process.  
There are several depolarization models: the \textit{Slab} model (\cite{Burn66}), the \textit{Tribble} model (\cite{Tribble91}), the \textit{Rossetti-Mantovani} model (\cite{Rossetti08}; \cite{Man09}) and the \textit{Repolarizer} model (\cite{Homan02}; \cite{Man09}, \cite{Hovatta12}). Many of them assume an optically thin emitting region and all of them make the assumption that we detect the same emitting region at each frequency. However, we see different synchrotron components in the most of our targets, thus our unresolved sources can have several overlaps of optically thick and optically thin components together that can modify the polarization behaviour from the simplified way described from the models. For simplicity we use three models that are just a mathematical generalisation of the various physical models (like in \cite{Farnes14}): a Gaussian, a power law and a Gaussian with a constant term. 
Some of the sources ($\sim$ 30\% of the sample; see the source 0845+0439 in Fig. 2 as an example) show a complex behaviour with a rising and decaying of the fractional polarization; for them none of the three models fit properly the data. We did not find any correlation between the depolarization models and the SEDs shapes or the RM values.

\section{Extreme in-band RM: the JVLA observations}
Wide band observations at L (1 GHz BW), S (2 GHz BW), C (4 GHz BW) and X (4 GHz BW) bands using the JVLA have been performed on the most extreme objects, i.e. the 12 sources showing a $|$RM$_{obs}$$|$$\geq$ 500 rad/m$^2$.
In Fig.5 we show an example of our preliminary results. Now, thanks to the high spectral resolution in frequency, we are able to follow the behaviour of the fractional polarisation and the polarisation angle. We can now better fit the depolarization behaviour.

Preliminary results from the JVLA data confirm the previous Effelsberg results that the RM of some sources deviates from the linear fit. Indeed, in the example in Fig.5d, we see a dramatic change of the polarization angle within the wideband. This could be an indication of several Faraday screens that differently rotate the polarisation angle, thus a proof of the complexity of the target. In order to understand this unusual behaviour, we will apply the Rotation measure synthesis, a powerful method, not affected by the n$\pi$ ambiguities, that can reconstruct the several Faraday depth components within the source (\cite{BrBr05}), thus mapping the ambient medium of the targets.

Moreover, in a few cases we noticed clear signs of variability (see e.g. Fig. 5). In this specific case, it could be justified because of the GPS nature of the target itself. This peculiar class of objects are characterised by very compact sources that are supposed to be young radio sources (\cite{O'Dea98}), thus probably characterised by some intense and periodic activity. In order to avoid variability effects, our JVLA observations were performed simultaneously at all frequencies.

\begin{figure}[]
\begin{center}
\includegraphics[width=0.48\textwidth]{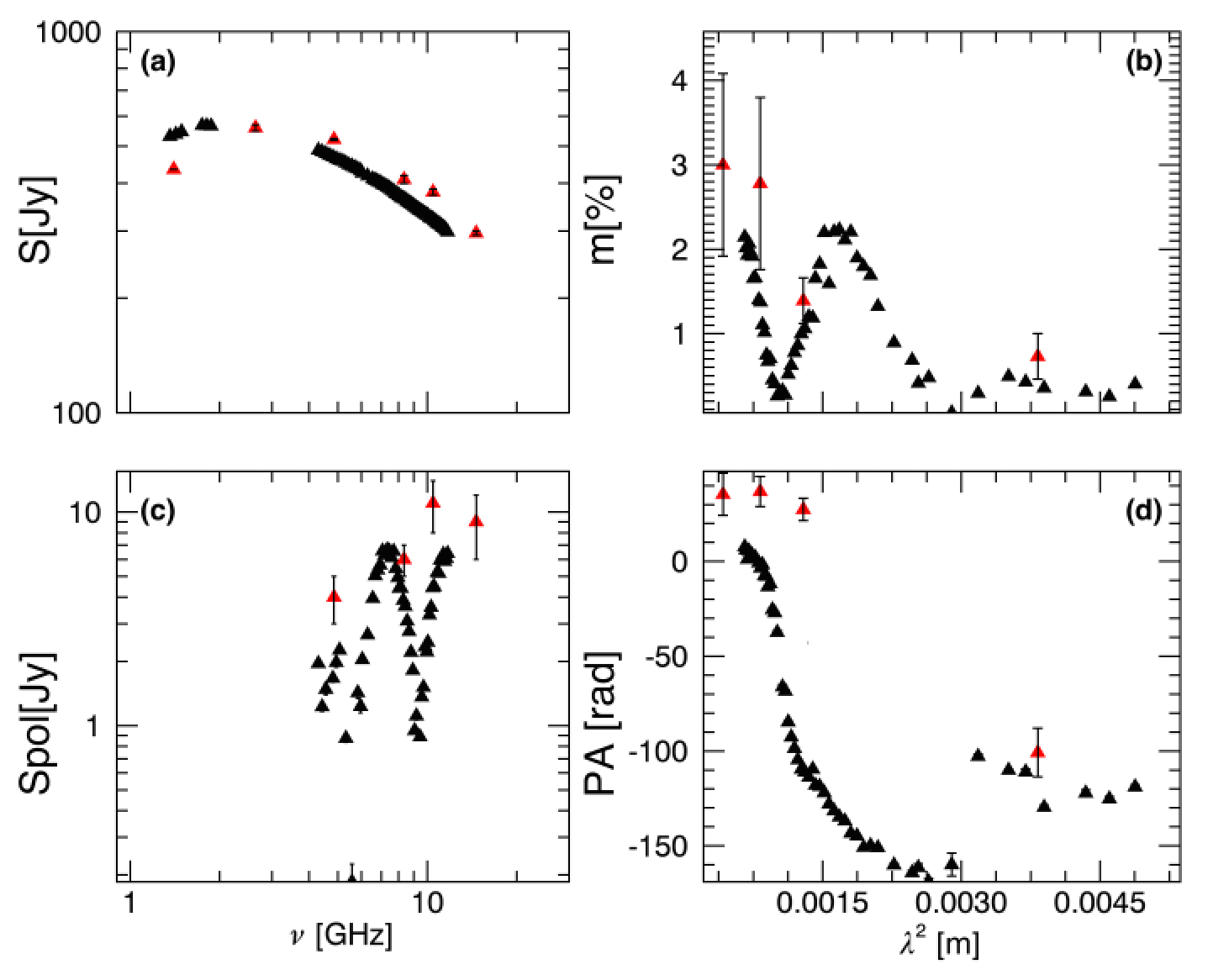}
\caption{Example of the JVLA data. The black triangles are JVLA data and the red triangle are the Effelsberg data.}
\label{default}
\end{center}
\end{figure}

\section{Conclusions}
With our observations we were able to find a number of sources with a potentially very high-RM. In some case, the observed $|$RM$|$ is higher than 1000 rad/m$^2$.
From the preliminary results of our project, we can conclude that:

\begin{itemize}
\item The high-RM candidates are not characterised by a specific object type. They are present among different object types: steep, flat and inverted radio spectrum sources. 

\item High-RM candidates are mainly sources with compact high frequency components, probably new growing radio components, thus objects in a particular compact young phase as the GPS and HFP sources or in a reactivated activity phase.
 
\item For some sources the RM deviates from a linear behaviour with $\lambda^2$. The well sampled frequency  data obtained with the JVLA, confirm this result. We interpret this as an indication of several Faraday screens within the medium.

\item A correlation between sources showing a restarting activity phase in their SEDs and a high-RM value has been found. We suggest that these sources are characterised by a really dense and/or a magnetised medium.

\item The behaviour of the fractional polarization \textit{m} with $\lambda^{2}$ have been fitted using simpler mathematical representations of the main physical depolarization models but no correlation between the SED type and the fractional polarization behaviour has been found. 

\end{itemize}
As a future work, we will analyse the already available data taken with high angular resolution VLBI technique using the European VLBI Network (EVN), at C and X bands, and the Very Long Baseline Array (VLBA), at U and K bands. 
We expect that these higher angular resolution observations will allow us to produce detailed polarisation maps together with maps of the polarisation angle of the different components. In this way, we will be able to identify the different contributions to the RM on the pc scale. Moreover, these observations will allow us to make good estimations of the magnetic field, one of the quantities that could contribute to the high RM.

\begin{acknowledgements}
Based on observations with the 100-m telescope of the MPIfR (Max-Planck-Institut f\"ur Radioastronomie) at Effelsberg.
The research leading to these results has received funding from
the  European Commission Seventh Framework Programme (FP/2007-2013)
under grant agreement No 283393 (RadioNet3).
AP is a member of the International Max Planck Research School (IMPRS) for Astronomy and Astrophysics
at the Universities of Bonn and Cologne.
C.C-G. acknowledges support by UNAM-DGAPA-PAPIIT grant number IA101214.
We thank the useful help from the Effelsberg operators.
\end{acknowledgements}


\begin{thebibliography}{100}
\expandafter\ifx\csname natexlab\endcsname\relax\def\natexlab#1{#1}\fi

\bibitem[{{Attridge} {et al.} 2005}]{Attridge05} Attridge, J.~M., Wardle, J.~F.~C., Homan, D.~C., \& Phillips, R.~B.\ 2005, Future Directions in High Resolution Astronomy, 340, 171 

\bibitem[{{Benn} {et al.} 2005 }]{Benn05} Benn, C.~R., Carballo, R.,  Holt, J., et al.\ 2005, \mnras, 360, 1455 

\bibitem[{{Brentjens} \& {de Bruyn} 2005}]{BrBr05} Brentjens, M.~A., \& de Bruyn, A.~G.\ 2005, A\&A, 441, 1217 

\bibitem[{{Burn} 1966}]{Burn66} Burn, B.~J.\ 1966, \mnras, 133, 67 

\bibitem[{{Condon} {et al.} 1998}]{Condon98} Condon, J.~J., Cotton, W.~D., Greisen, E.~W., et al.\ 1998, \aj, 115, 1693 

\bibitem[{{Czerny} {et al.} 2009}]{Cz09}
{Czerny}, B. and {Siemiginowska}, A. and {Janiuk}, A. and {Nikiel-Wroczy{\'n}ski}, B. and {Stawarz}, {\L}., 2009, \apj, 698, 840-851

\bibitem[{{Dallacasa} {et al.} 2000}]{Dcasa00} {Dallacasa}, D. and {Stanghellini}, C. and {Centonza}, M. and {Fanti}, R., 2000, A\&A, 363, 887-900

\bibitem[{{Fanti} {et al.} 1990}]{Fanti90}
{Fanti}, R. and {Fanti}, C. and {Schilizzi}, R.~T. and {Spencer}, R.~E. and {Nan Rendong} and {Parma}, P. and {van Breugel}, W.~J.~M. and {Venturi}, T., 1990,  A\&A, 231, 333-346

\bibitem[{{Fanti} {et al.} 2004}]{Fanti04} Fanti, C., Branchesi, M., Cotton, W.~D., et al.\ 2004, A\&A, 427, 465

\bibitem[{{Farnes} {et al.} 2014}]{Farnes14} Farnes, J.~S., Gaensler, B.~M., \& Carretti, E.\ 2014, \apjs, 212, 15 

\bibitem[{{Gopal-Krishna} {et al.} 1983}]{Gopal-Krishna83}
{Gopal-Krishna} and {Patnaik}, A.~R. and {Steppe}, H., 1983, A\&A, 123, 107-110

\bibitem[{{Homan} {et al.} 2002}]{Homan02} Homan, D.~C., Ojha, R., Wardle, J.~F.~C., et al.\ 2002, \apj, 568, 99 

\bibitem[{{Hovatta} {et al.} 2012}]{Hovatta12} Hovatta, T., Lister, M.~L., Aller, M.~F., et al.\ 2012, \aj, 144, 105 

\bibitem[{{Jorstad} {et al.} 2007 }]{Jorstad07} Jorstad, S.~G., Marscher, A.~P., Stevens, J.~A., et al.\ 2007, \aj, 134, 799 

\bibitem[{{Kato} {et al.} 1987}]{Kato87} Kato, T., Tabara, H., Inoue, M., \& Aizu, K.\ 1987, Nature, 329, 223 

\bibitem[{{Kravchenko} {et al.} 2015}]{Kravchenko15} Kravchenko, E.~V., Cotton, W.~D., \& Kovalev, Y.~Y.\ 2015, IAU Symposium, 313, 128 

\bibitem[{{Mantovani} {et al.} 2009}]{Man09} Mantovani, F., Mack, K.-H., Montenegro-Montes, F.~M., Rossetti, A., \& Kraus, A.\ 2009, A\&A, 502, 61 

\bibitem[{{Marecki} {et al.} 2006}]{Ma06}
{Marecki}, A. and {Thomasson}, P. and {Mack}, K.-H. and {Kunert-Bajraszewska}, M., 2006, A\&A, 408, 479-487

\bibitem[{{O'Dea} 1998}]{O'Dea98} O'Dea, C.~P.\ 1998, IAU 
Colloq.~164: Radio Emission from Galactic and Extragalactic Compact 
Sources, 144, 291 

\bibitem[{{O'Dea} {et al.} 1991}]{O'Dea91}
{O'Dea}, C.~P. and {Baum}, S.~A. and {Stanghellini}, C., 1991, \apj, 380, 66-77

\bibitem[{{Orr} \& {Browne} 1982}]{OB82}
{Orr}, M.~J.~L. \& {Browne}, I.~W.~A., 1982, \mnras, 200, 1067-1080

\bibitem[{{Rossetti} {et al.} 2008}]{Rossetti08} Rossetti, A., Dallacasa, D., Fanti, C., Fanti, R., \& Mack, K.-H.\ 2008, A\&A, 487, 865 

\bibitem[{{Rossetti} {et al.} 2009}]{RosMan09} Rossetti, A., Mantovani, F., Dallacasa, D., et al.\ 2009, A\&A, 504, 741 

\bibitem[{{Saikia} 1988}]{Saikia88}
{Saikia}, D.~J., 1988, Lecture Notes in Physics, 307, 317

\bibitem[{{Saikia} \& {Salter} 1988}]{SaSa88} Saikia, D.~J., \& Salter, C.~J.\ 1988, \araa, 26, 93 

\bibitem[{{Saikia} \& {Jamrozy} 2009}]{Saikia09}
{Saikia}, D.~J. and {Jamrozy}, M., 2009, Bulletin of the Astronomical Society of India, 37, 63-89

\bibitem[{{Tribble} 1991}]{Tribble91} Tribble, P.~C.\ 1991, \mnras, 250, 726 

\bibitem[{{Trippe} {et al.} 2012}]{Trippe12} Trippe, S., Bremer, M., Krichbaum, T.~P., et al.\ 2012, \mnras, 425, 1192 

\bibitem[{{Urry} \& {Padovani} 1995}]{UP95}
{Urry}, C.~M.  \& {Padovani}, P., 1995, \pasp, 107, 803

\bibitem[{{Vallee} 1980}]{Val80} Vallee, J.~P.\ 1980, A\&A, 86, 251 

\bibitem[{{White} {et al.} 1997}]{White97} White, R.~L., Becker, R.~H., Helfand, D.~J., \& Gregg, M.~D.\ 1997, \apj, 475, 479 

\bibitem[{{Zavala} \& {Taylor} 2004}]{Zavala04} Zavala, R.~T., \& Taylor, G.~B.\ 2004, \apj, 612, 749 

\end{thebibliography}
\end{document}